\documentclass[final,3p,times,twocolumn]{elsarticle-mod}

\usepackage{amssymb}

\usepackage[hyphens]{url}

\usepackage{amsmath}

\usepackage{color}
\usepackage{xtab}
\usepackage{overpic}
\usepackage[para]{threeparttable}
\usepackage{etoolbox}
\usepackage{pict2e}

\usepackage{fancyhdr}

\appto\TPTnoteSettings{\footnotesize}

\journal{NSS Space Settlement Journal}

\begin{document}

\begin{frontmatter}



\title{Control of habitat's carbon dioxide level by biomass burning}


\author[FMI]{Pekka Janhunen\corref{cor1}}
\ead{pekka.janhunen@fmi.fi}
\ead[url]{http://www.electric-sailing.fi}

\address[FMI]{Finnish Meteorological Institute, Helsinki, Finland}
\cortext[cor1]{Corresponding author}

\begin{abstract}
  Consider a free-space settlement with a closed ecosystem. Controlling the habitat's carbon dioxide level is a nontrivial problem
  because the atmospheric carbon buffer per biosphere area is
  smaller than on Earth. Here we show that the problem can be solved by
  burning agricultural waste.  Waste biomass is stored and dried, and
  burned whenever plant growth has lowered the atmospheric carbon dioxide level so that replenishment is needed.
  The method is robust, low-tech and scalable.
  The method also leaves the partial pressure of oxygen unchanged.  In the initial
  growth phase of the biosphere, one can obtain the carbon dioxide by
  burning sugar or carbon, which can be sourced from carbonaceous
  asteroid materials. This makes it possible to bootstrap the biosphere
  without massive biomass imports from Earth.
\end{abstract}


\end{frontmatter}





\section{Introduction}

\begin{figure}
\includegraphics[width=0.49\textwidth]{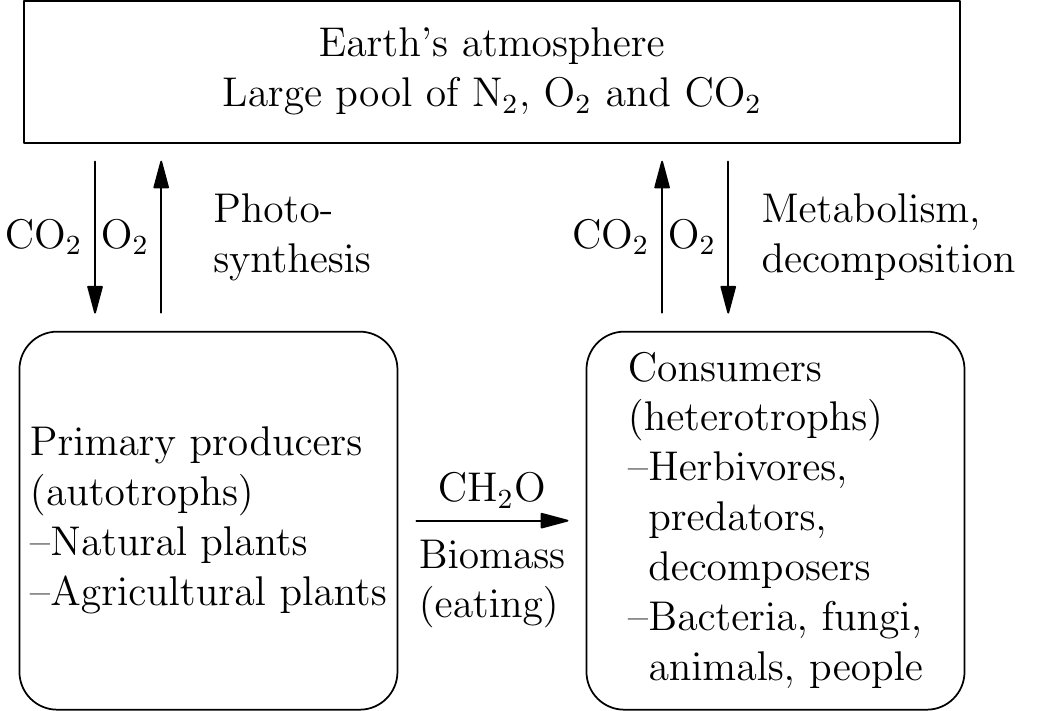}
\caption{Carbon cycle on Earth.}
\label{fig:earth}
\end{figure}

Space settlements need a nearly closed ecosystem for food production. One of
the fundamental parts of a closed ecosystem is the carbon cycle. In
the carbon cycle (Fig.~\ref{fig:earth}), plants fix carbon from
atmospheric CO$_2$ by photosynthesis, producing biomass [approximately
sugar, net formula \textit{n}(CH$_2$O)] and liberating oxygen,
\begin{equation}
\mathrm{CO}_2 + \mathrm{H}_2\mathrm{O} + \mathrm{light} \to \mathrm{CH}_2\mathrm{O} + \mathrm{O}_2.
\label{eq:photosynthesis}
\end{equation}
The biomass is consumed and metabolised by decomposers, animals and
people. Metabolism is the reverse reaction of photosynthesis,
\begin{equation}
\mathrm{CH}_2\mathrm{O} + \mathrm{O}_2 \to \mathrm{CO}_2 +
\mathrm{H}_2\mathrm{O} + \mathrm{energy}.
\label{eq:burning}
\end{equation}

On Earth, the atmospheric CO$_2$ and the biospheric CH$_2$O contain
comparable amounts of carbon. This is so because the amount of carbon in the atmospheric CO$_2$
is 1.66 kgC/m$^2$, while the world average biospheric carbon is 1.08
kgC/m$^2$ \citep{Bar-OnEtAl2018}\footnote{550 billion tonnes of
  carbon\citep[Table 1]{Bar-OnEtAl2018} is 1.08 kgC/m$^2$.}.
Because the atmospheric carbon buffer is large, on Earth the atmospheric CO$_2$ level is not sensitive to
fluctuations in the primary production of the biosphere.
The Earth's atmosphere is massive (10 tonnes per square metre),
while most of Earth's surface area is open ocean, desert or glacier
so that the globally averaged biomass
  areal density is only moderate.  For example in
average African tropical rainforest, the
carbon stock is 18.3 kgC/m$^2$ i.e.~183 Mg/ha \citep[Table 2]{SullivanEtAl2017},
which is as much as 17 times larger than the global average.

In a space settlement, the atmosphere mass is likely to be much less
than 10 tonnes/m$^2$. In O'Neill's original large habitat concepts
\citep{ONeill1974,ONeill1977}, the atmosphere had several kilometres
depth. However, a massive atmosphere includes a lot of
nitrogen. Nitrogen is not too abundant on asteroids, and would only be
widely available in the outer solar system. One way to avoid the
nitrogen supply problem would be to use a reduced pressure pure oxygen
atmosphere, but then the risk of fire would be increased since the
flame is not cooled by inert gas. Also birds and insects (needed for
pollination) would have difficulty in flying in a pure oxygen
atmosphere, because its mass density would be several times less than
on Earth. Hence it is likely that most settlements would prefer to use
a shallower N$_2$/O$_2$ atmosphere of e.g.~$\sim 50$ m depth
\citep{Janhunen2018}. A 50 m height allows forests with maximum tree
height of $\sim 30$ m plus some room for horizontal winds to mix gases
above the treetops. The nitrogen (47 kg/m$^2$) can be obtained
from the asteroids, as a byproduct of the mining that produces
the combined structures and radiation shielding of the settlement ($10^4$ kg/m$^2$).

Carbon dioxide is necessary for plants to grow. To maintain good
growth, the concentration should be at least $\sim 300$ ppmv (parts
per million by volume). The pre-industrial level on Earth
was 280 ppmv, which, as we know, already allowed plants to grow reasonably well. On the
other hand, for human safety the amount should not exceed $\sim 2000$
ppmv. The U.S.~occupational safety limit for a full working day is
5000 ppmv. The atmospheric concentration must be clearly less, however,
since local concentration near sources is always higher than the
atmospheric one. An example of local source is indoors where people continuously produce CO$_2$ by breathing.

A shallow atmosphere is unable to absorb fluctuations in the biomass
carbon pool while keeping the CO$_2$ level within safe bounds. The
timescales can be rather fast. A tropical rainforest can bind 2.0
kgC/m$^2$/year \citep{wikipediaBiomass,RicklefsAndMiller2000}, so in a
shallow 50 m atmosphere, maximal plant growth could reduce the
concentration of CO$_2$ by 1000 ppmv in as short time as 4.5 days. In
temperate forest the rate of biomass production is somewhat less (1.25
kgC/m$^2$/year) and in cultivated areas even less (0.65
kgC/m$^2$/year)\citep{wikipediaBiomass,RicklefsAndMiller2000}, but the
timescales are still only weeks. Hence the atmospheric CO$_2$ must be
controlled by technical means, which is the topic of this paper.


\rfoot{\textit{NSS Space Settlement Journal}}
\pagestyle{fancy}

\section{Feasibility of a closed ecosystem}
\label{sect:feasibility}

There are many examples of nearly semi-closed small ecosystems that
interact with the rest of Earth's biosphere mainly via air only:
a potted flower, a vivarium, a fenced garden, a small island, etc. To
turn a semi-closed system into a fully closed one, one only needs to
worry about a few gases. This
is an engineering task, where the complexity of biology has been factored out. More specifically, there are five
parameters to consider:
\begin{enumerate}
\item O$_2$ partial pressure. Oxygen is needed for humans and animals to breath, and the partial pressure should be about 0.21 bar.
\item N$_2$ partial pressure. Nitrogen is needed for fire safety and for birds and insects to fly, and the partial pressure should be about 0.79 bar.\footnote{We do not consider argon and other noble gases because they are even less abundant on asteroids than N$_2$. Also, at high concentrations some noble gases have narcotic effects.}
\item CO$_2$ concentration. Carbon dioxide is needed by plants to
  grow, but too high a value is unsafe to people. The allowed range is 300--2000 ppmv.
\item CH$_4$ concentration. Methane is not needed so the lower limit is zero, but if generated by the biosphere, it is tolerable up to 30 mbar, which is well below the ignition limit of 44 mbar. Methane's only health effect is oxygen displacement, which is however negligible at 0.03 bar.
\item Other gases should remain at low concentration.
\end{enumerate}

Considering oxygen, a biosphere does not fix it from the
atmosphere. The oxygen atoms that biomass CH$_2$O contains originate
from the water that enters photosynthesis. When organisms
do metabolism and breathe (Eq.~\ref{eq:burning}), they transform O$_2$ molecules into
CO$_2$ molecules, but the process involves no net transfer of O atoms
from the atmosphere into the body. Hence one does not need to do
anything special to maintain the right O$_2$ partial
pressure.


Considering N$_2$, a biosphere fixes some of it since nitrogen is a
key nutrient, present in proteins and DNA. The C:N ratio of cropland
soil is 13.2 and for other biomes it varies between 10.1 and
30 \citep[Table 1]{WangEtAl2010}. For leaves, wood and roots the C:N
ratio is higher \citep{WangEtAl2010}. To get an upper limit, the
carbon stock of average African rainforest is 18.3 kgC/m$^2$
\citep{SullivanEtAl2017}. With the minimal soil C:N ratio across
biomes of 10.1, this corresponds to 1.81 kgN/m$^2$ of fixed
nitrogen. But the mass of nitrogen in a 50 m high atmosphere is 46
kgN/m$^2$, so clearly the biosphere can assimilate only a small
fraction of atmospheric N$_2$.  Hence one does not need to do anything
special with N$_2$, either. Its partial pressure will remain
sufficiently close to the initial value. Circulation of
  nitrogen from the point of view of nutrient supply is a related
  topic \citep{JewellAndValentine2010}, which is however outside
  the scope of this paper.

Thus, since N$_2$ and O$_2$ are not changed too much by the
biosphere, the task of maintaining a good atmosphere
is reduced to three issues:
\begin{enumerate}
\item Maintaining CO$_2$ within the 300--2000 ppmv bounds. This is treated
  in the next section.
\item Ensuring that if net methane is emitted by the biosphere, its
  concentration does not increase beyond $\sim 3$\,\% by
  volume. \footnote{On Earth the methane concentration is 1.8 ppmv, which is responsible for part of the terrestrial greenhouse effect. For atmospheric height of 50 m, a similar greenhouse effect arises at 200 times higher concentration, i.e.~at 360 ppmv. Thus a 3\,\% (30,000 ppmv) methane concentration would cause a significant greenhouse effect for a 50 m atmosphere, which should be taken into account in the settlement's heat budget. Greenhouse effects are nonlinear so quantitative prediction would need modelling.}
\item Ensuring that the concentration of other gases stay low. This may
  possibly happen automatically, because plants are known to
  remove impurities from air \citep{WolvertonEtAl1989}. We shall say a bit more on this in the Discussion section below.
\end{enumerate}

\section{Biomass burning}

Above we described the carbon cycle problem of the orbital space
settlement. The problem is that the settlement's atmosphere is much
shallower than on Earth, and hence the atmospheric carbon buffer is
much smaller than the biospheric carbon stock. Fluctuations in the
amount of biospheric carbon can occur for many reasons, and the
fluctuations would cause the atmospheric CO$_2$ concentration to go off bounds.

\begin{figure*}
\centering
\includegraphics[width=0.7\textwidth]{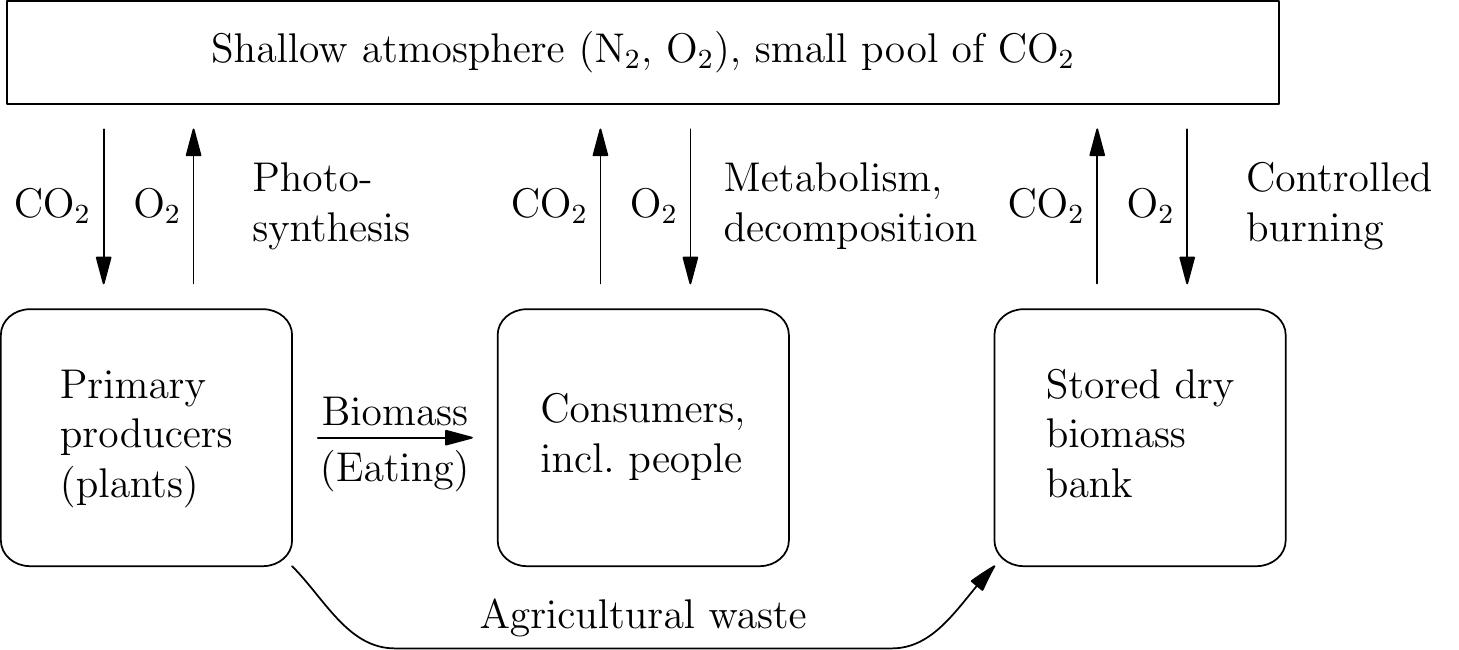}
\caption{Carbon cycle in the settlement.}
\label{fig:carbcycle}
\end{figure*}

A way to solve the problem is to store some biomass and to burn it
when the atmosphere needs more CO$_2$
(Fig.~\ref{fig:carbcycle}). Agricultural waste is a necessary
byproduct of food production. One stores the waste biomass in such a
way that it does not decompose and then burns it at a controlled
rate. Methods to store biomass include drying, freezing and
freeze-drying. Drying is feasible at least if the relative humidity
is not too high.

It is sufficient for only part of the biomass to go through the
storing and burning pathway. The higher the burned fraction is, the
larger is the CO$_2$ control authority of the scheme. The control
authority is sufficient if the total amount of carbon in the
settlement exceeds the maximum mass of carbon that can be fixed in
living organisms at any one time. When the atmospheric CO$_2$ drops
below a target value, one burns some stored biomass.  If there is too
much of CO$_2$ in the atmosphere, one ceases the burning activity for
a while. After some delay plant growth will take down the CO$_2$
concentration.

Burning consumes oxygen, but the same amount of oxygen is liberated
into the atmosphere when the CO$_2$ is used by photosynthesis
(Eqs.~\ref{eq:photosynthesis} and \ref{eq:burning}). Thus the O$_2$
concentration stays constant, apart from an insignificant
  part that exists temporarily as CO$_2$. This is especially advantageous in the
build-up phase of the biosphere. In the build-up phase, one needs to
add carbon constantly to the atmosphere, as trees and other plants are
growing. Depending on the type of ecosystem we are building,
  the growth phase might last up to tens or even hundreds of years as
  trees grow and the soil builds up. It is not necessary to wait for
  the growth phase to finish until people can move in, but while the
  growth phase is ongoing, one must be
  prepared to put in new carbon as needed to avoid CO$_2$ starvation. If this carbon would be added in the form of new CO$_2$ from
an external tank, for example, the level of atmospheric oxygen would
build up. However, if one adds the carbon by burning biomass, sugar or
carbon, the O$_2$ level stays constant.\footnote{Burning hydrocarbons
  ($\sim$CH$_2$) in the buildup phase is not recommended, because then
  net consumption of O$_2$ would take place as oxygen would be bound
  with hydrogen to make water.} Carbon can be sourced from
carbonaceous asteroids. Possibly sugar [net formula \textit{n}(CH$_2$O)] can be synthesised from
C-type asteroids as well. Thus the biosphere can be bootstrapped
without massive importing of biomass from Earth.

When burning biomass, the rate must be controllable and fire safety
must be maintained.  One also wants to minimize smoke production (particulate
emission), because otherwise the settlement's sunlight-passing windows
would need frequent washing and because we want to avoid
  atmospheric pollution \citep{SoilleuxAndGunn2018}. One way to facilitate clean burning is to
mechanically process the biomass (or part of it which is used in the
ignition phase) into some standardised form such as pellets
\citep{ThomsonAndLiddell2015} or wood chips. It is also possible to use a
bioreactor to turn the biomass into biogas (methane) which burns
without smoke. To further reduce smoke, one might add an electrostatic smoke precipitator in
  the smokestack.
A combination of approaches is also possible. One can
ignite the flame using easy fuel and then continue with more
unprocessed material. The burning activity could be continuous, but in a
50 m high atmosphere, enough constant CO$_2$ is reached by a
daily burning session.

Atmospheric pollution should be avoided, so smoke production
  should be minimised. However,
  plants and soil are known to clean up the atmosphere rather well
  \citep{WolvertonEtAl1989}. Hopefully, if
  the above measures to promote clean burning are used, the plants can
  accomplish the rest so that the atmosphere remains clean. To investigate the question
  experimentally, one could burn
  biomass inside a greenhouse by different methods, while using
  standard air quality monitoring equipment for measuring the atmosphere.

In a rainforest, the maximum carbon fixation rate is 2
  kgC/m$^2$/year and in a cultivated area it is 0.65 kgC/m$^2$/year (see
  last paragraph of Introduction). If the average is $\sim 1$
  kgC/m$^2$/year and if 50\,\% of it is burned while the remaining
  part is decomposed naturally or eaten as crop, then the burned
  amount is 0.5 kgC/m$^2$/year, which corresponds to 34 kg of dry
  biomass per hectare per day. When wood is burned, the mass fraction
  of ash varies between 0.43 and 1.8 per cent \citep[Table
  1]{MisraEtAl1993}, so that the ash produced is
  a few hundred grams per day per hectare. The ash must be
  distributed evenly back into the environment. The amount of ash is
  modest enough that the settlers could even do the spreading manually
  if they wish. The heat produced by the
  burning is of the order of 0.8 W/m$^2$ as a temporal average, which
  is two orders of magnitude less than the heat dissipation of sunlight, or
  artificial light if that is employed.

  In reality, a smaller burning rate than
  this calculation would probably suffice. It is only necessary to
  burn enough biomass to maintain sufficient control authority of the
  CO$_2$ level. Burning as much 50\,\% of the growth is likely to
  be overkill, but we assume it to arrive at a conservative estimate.

Animal and human wastes are not burned, but composted to make
  leaf mold which is spread onto the fields. Our recommendation is to
  primarily burn agricultural plant waste which is poor in non-CHO
  elements, comprising substances such as cellulose, lignin and
  starch. In this way we avoid unnecessarily releasing fixed nitrogen
  and other valuable nutrients into the atmosphere, where they would
  also be pollutants.

\section{Backup techniques}

As was pointed out above, typically the biosphere is not able to fix
so much oxygen or nitrogen that it would change the atmospheric
concentrations of these gases too much. However, to facilitate dealing
with accident scenarios like air leakages or atmospheric poisonings,
having compressed or liquefied O$_2$ and N$_2$ available could be
desirable \footnote{In addition, one probably wants to divide the
  settlement into separately pressurisable sectors
  \citep{Janhunen2018} so that people can be evacuated from a sector
  that suffered an accident.}.  If so, it may make sense to also have
a mechanism available for moving O$_2$ and N$_2$ selectively from the
habitat into the tanks by e.g.~cryogenic distillation of air \citep{SoilleuxAndGunn2018}. If such
process is implemented, then CO$_2$ is also separable. For managing
CO$_2$, such process would be energetically inefficient because
e.g.~to reduce the CO$_2$ concentration into one half, one has to
process 50\,\% of the air by liquefaction, separating out the CO$_2$
and returning the O$_2$ and N$_2$ back into the settlement. However,
if energy is available, energy efficiency is not a requirement for
backup strategies.  Chemical scrubbing of CO$_2$ into amines or
hydroxides is another possible backup strategy for emergency removal
of CO$_2$. Table \ref{tab:alternatives} lists these
  alternatives and their potential issues.




\begin{table}[htb]
\caption{Some alternatives of habitat CO$_2$ control.}
\label{tab:alternatives}
\begin{tabular}{|l|l|}
\hline
\textbf{Method}  & \textbf{Potential issues} \\
\hline
Biomass burning & --Smoke \\
& --Need to handle fire \\
\hline
Cryo-distillation & --Power-intensive \\
or air & --Reliability concern/moving parts \\
& --Mass overhead of CO$_2$ tanks \\
\hline
Scrubbing & --Reliability concern/moving parts\\
into amines & --Safety concern due to chemicals \\
or hydroxides &  \\
\hline
\end{tabular}
\end{table}

\section{Discussion}

As described in Section \ref{sect:feasibility}, gardens,
vivariums and other widespread examples of semi-closed (i.e., only
  gases exchanged) ecosystems show that closed biospheres are
feasible. The only issue is to maintain the right atmospheric
composition, but this is only a technical problem to which there are
many solutions. The biomass burning is one of them. The complexity of
biology cannot spoil the feasibility of closed biospheres. If it
could, it would already have been seen in gardens and vivariums. The
complexity of biology is factored out of the feasibility equation.

The biomass burning method works, as such, only in a tropical climate
with no dark season. During dark season photosynthesis is stopped
and the level of CO2 would probably build up too high in the
atmosphere. Therefore, if seasons are wanted, one has to use sectoring
such as discussed in \citet{Janhunen2018}. Different sectors must
then be phased in different seasons and air must be exchanged between
sectors.

Biomass burning seems to be a straightforward, scalable, low-tech
and reliable solution. A possible drawback is the production of smoke.
As on Earth, plants and soil are absorbers of air pollution, but production
of smoke should nevertheless be minimised to prevent health issues.
In addition, smoke in a settlement environment is more harmful than on Earth,
because the settlement has windows through which sunlight enters, or
if it is artificially lighted, the lamps have cover glasses.
The production of smoke can be minimised by technical means such as
igniting the fire by a biogas flame or by mechanically making the
biomass into pellets or other granular form.

Biomass burning involves fire, and fire is in principle a risk because
conflagration in a space settlement would be very dangerous.
Concerning fire risk in general, it is not feasible to eliminate
it entirely by removing all possible ignition sources, e.g.~because
electric equipment is necessary and malfunctioning electric
equipment is a potential ignition source. The risk of wildfire can be
lowered by having frequent artificial rain so that the environment is
fresh and green. Lush nature also boosts agricultural output and is
good for aesthetic reasons. However, not everything can be humid since
the stored biomass must be dry in order to burn cleanly. Thus
the relative humidity should be less than 100\,\%, which is
also convenient for people.  To reduce the fire risk further, an easy
way is to store the dry biomass far from the locations where it is
burned.  Artificial rain or sprinkler system must
be possible to turn on quickly in case a fire breaks out.

Also other approaches for reducing the fire risk are possible. For
example, one can freeze-dry the biomass and store it in a refrigerated
space. Storage under nitrogen-enriched atmosphere is another
possibility, which eliminates the fire risk during
storage. Nitrogen-enriched gas can be made e.g.~by filtering air
through certain polymeric membranes.

The methods discussed in this paper do not involve moving materials
through airlocks. Thus there is no issue of losing atmospheric gases
into space.

After O$_2$, N$_2$ and CO$_2$ are controlled, the remaining issue is
how to keep the level of other volatile compounds low. Plants remove
harmful impurities \citep{WolvertonEtAl1989}, but they also produce some volatile organic
compounds (VOCs) of their own, such as isoprene and terpenes. This
smell of plants can be experienced e.g.~in greenhouses and it is
generally considered pleasant. However, too much of a good thing is potentially
a bad thing, so let us briefly discuss loss mechanisms of VOCs. It is
thought that the hydroxyl radical OH is an important ``detergent'' of
the troposphere that oxidises VOCs \citep{LelieveldEtAl2004}. On
Earth, the primary formation of OH is by solar UV
and is highest in the tropics where the solar zenith angle is
smallest, the stratospheric ozone layer is thinnest and the humidity
is highest \citep{LelieveldEtAl2004}. Thus, in the settlement it might
be a good idea not to filter out the solar UV entirely, but let a
small part of it enter so that the UV radiation level mimics the
conditions in Earth's troposphere, thus maintaining some OH
to remove VOCs and also methane by oxidation.

One of the referees pointed out that the carbon
stock of soil might potentially grow in time due to incomplete
decomposing. While certain biomes like some wet peatlands
exhibit slow continuous carbon accumulation, typical biomes such as forests
have moderate carbon stocks \citep{WangEtAl2010} that
presumably have not essentially grown even in
millions of years. Earth's significant fossil coal deposits are thought to
have been accumulated before lignin-degrading organisms developed
around the end of the Carboniferous period
\citep{FloudasEtAl2012}. On modern Earth, termites are good
lignin decomposers \citep{ButlerAndBuckerfield1979} so their presence in the habitat ecosystem
could be beneficial for efficient carbon circulation.

\section{Summary and conclusions}

Controlling a habitat's carbon dioxide level is a nontrivial problem
because the atmospheric volume per biosphere area is typically much
smaller than on Earth. The problem is important because too low CO$_2$
($\lessapprox 300$ ppmv) slows down plant growth and thus food
production while too high concentration ($\gtrapprox 2000$ ppmv) begins
to cause health problems for people.

The problem can be solved by biomass burning. In particular,
agricultural waste is a necessary byproduct of food production. One
can dry and store this biomass and burn some of it when the CO$_2$
level in the settlement's atmosphere drops too low. The method is
straightforward, robust and low-tech. It ensures large control
authority of the CO$_2$ while keeping the O$_2$ partial pressure
unchanged. The method scales to habitats of all sizes.

In the initial growth phase of the biosphere, one can obtain the
CO$_2$ by burning sugar or carbon. They can be sourced from
carbonaceous asteroid materials so that bootstrapping the biosphere
does not require lifting large masses from Earth.

Closed ecosystems in habitats are feasible. We know this because there
are many examples of semi-closed ecosystems such as gardens --
  and because it has been done e.g.~in Biosphere-II and BIOS-1, 2 and 3\citep{SalisburyEtAl1997}. 
Maintaining the atmosphere is an engineering problem that can
be solved. For gases other than CO$_2$, the problem is in fact
solved automatically. For the control of CO$_2$, the biomass burning
method seems simple and effective.

\section{Acknowledgement}

The results presented have been achieved under the framework of the
Finnish Centre of Excellence in Research of Sustainable Space (Academy
of Finland grant number 312356). I am grateful to journalist Hanna
Nikkanen for providing her compilation of papers on the topic.







\end{document}